\documentclass[aps,superscriptaddress,showpacs,floatfix,amsmath,amssymb,twocolumn]{revtex4}
\usepackage{epsfig}
\usepackage{dcolumn}
\usepackage{bm}

\usepackage{color}
\begin{document}

\title{Spectroscopy of $^{28}$Na: shell evolution toward the drip line}

\author{A.~Lepailleur}
\affiliation{Grand Acc\'el\'erateur National d'Ions Lourds (GANIL),
CEA/DSM - CNRS/IN2P3, B.\ P.\ 55027, F-14076 Caen Cedex 5, France}

\author{K.~Wimmer}
\affiliation{Department of Physics, The University of Tokyo, Hongo, Bunkyo-ku, Tokyo 113-0033, Japan}
\affiliation{Department of Physics, Central Michigan University, Mt. Pleasant, Michigan 48859, USA}
\affiliation{National Superconducting Cyclotron Laboratory, Michigan State University, East Lansing, Michigan 48824, USA}

 \author{A.~Mutschler}
\affiliation{Institut de Physique Nucl\'eaire, IN2P3-CNRS,
F-91406 Orsay Cedex, France}

\author{O.~Sorlin}
\affiliation{Grand Acc\'el\'erateur National d'Ions Lourds (GANIL),
CEA/DSM - CNRS/IN2P3, B.\ P.\ 55027, F-14076 Caen Cedex 5, France}

\author{V.~Bader}
\affiliation{Department of Physics and Astronomy and National Superconducting Cyclotron Laboratory, Michigan State University, East Lansing, Michigan, 48824-1321, USA}

\author{C.~Bancroft}
\author{D.~Barofsky}
\affiliation{Department of Physics, Central Michigan University, Mt. Pleasant, Michigan 48859, USA}

\author{B.~ Bastin}
\affiliation{Grand Acc\'el\'erateur National d'Ions Lourds (GANIL),
CEA/DSM - CNRS/IN2P3, B.\ P.\ 55027, F-14076 Caen Cedex 5, France}

\author{T.~Baugher}
\author{D. Bazin}
\affiliation{Department of Physics and Astronomy and National Superconducting Cyclotron Laboratory, Michigan State University, East Lansing, Michigan, 48824-1321, USA}

\author{V.~Bildstein}
\affiliation{Department of Physics, University of Guelph, Guelph, ON N1G 2W1, Canada}

\author{C.\ Borcea}
\affiliation{IFIN-HH, P. O. Box MG-6, 76900 Bucharest-Magurele, Romania}

\author{R.\ Borcea}
\affiliation{IFIN-HH, P. O. Box MG-6, 76900 Bucharest-Magurele, Romania}

\author{B. A. Brown}
\affiliation{Department of Physics and Astronomy and National Superconducting Cyclotron Laboratory, Michigan State University, East Lansing, Michigan, 48824-1321, USA}

\author{L.~Caceres}
\affiliation{Grand Acc\'el\'erateur National d'Ions Lourds (GANIL),
CEA/DSM - CNRS/IN2P3, B.\ P.\ 55027, F-14076 Caen Cedex 5, France}

\author{A. Gade}
\affiliation{Department of Physics and Astronomy and National Superconducting Cyclotron Laboratory, Michigan State University, East Lansing, Michigan, 48824-1321, USA}


\author{ L.\ Gaudefroy}
\affiliation{CEA, DAM, DIF, F-91297 Arpajon, France}

\author{ S.\ Gr\'evy}
\affiliation{Centre d`\'Etudes Nucl\'eaires de Bordeaux Gradignan-UMR 5797, CNRS/IN2P3, Universit\'e de Bordeaux 1, Chemin du Solarium, BP 120, 33175 Gradignan, France}

\author{G. F. ~ Grinyer}
\affiliation{Grand Acc\'el\'erateur National d'Ions Lourds (GANIL),
CEA/DSM - CNRS/IN2P3, B.\ P.\ 55027, F-14076 Caen Cedex 5, France}

\author{H. Iwasaki}
\affiliation{Department of Physics and Astronomy and National Superconducting Cyclotron Laboratory, Michigan State University, East Lansing, Michigan, 48824-1321, USA}

\author{E. Khan}
\affiliation{Institut de Physique Nucl\'eaire, IN2P3-CNRS,
  F-91406 Orsay Cedex, France}

\author{T.~Kr\"oll}
\affiliation{Institut f\"ur Kernphysik, Technische Universit\"at Darmstadt, 64289 Darmstadt, Germany}

\author{C.~Langer}
\affiliation{National Superconducting Cyclotron Laboratory, Michigan State University, East Lansing, Michigan 48824, USA}

\author{A.~ Lemasson}
\affiliation{Grand Acc\'el\'erateur National d'Ions Lourds (GANIL),
CEA/DSM - CNRS/IN2P3, B.\ P.\ 55027, F-14076 Caen Cedex 5, France}
\affiliation{Department of Physics and Astronomy and National Superconducting Cyclotron Laboratory, Michigan State University, East Lansing, Michigan, 48824-1321, USA}

\author{O. ~ Llidoo}
\affiliation{Grand Acc\'el\'erateur National d'Ions Lourds (GANIL),
CEA/DSM - CNRS/IN2P3, B.\ P.\ 55027, F-14076 Caen Cedex 5, France}

\author{J.~Lloyd}
\affiliation{Department of Physics, Central Michigan University, Mt. Pleasant, Michigan 48859, USA}

\author{F.\ Negoita}
\affiliation {IFIN-HH, P. O. Box MG-6, 76900
Bucharest-Magurele, Romania}

\author{F.~ de Oliveira Santos}
\affiliation{Grand Acc\'el\'erateur National d'Ions Lourds (GANIL),
CEA/DSM - CNRS/IN2P3, B.\ P.\ 55027, F-14076 Caen Cedex 5, France}

\author{G.~Perdikakis}
\affiliation{Department of Physics, Central Michigan University, Mt. Pleasant, Michigan 48859, USA}
\affiliation{National Superconducting Cyclotron Laboratory, Michigan State University, East Lansing, Michigan 48824, USA}

\author{F. Recchia}
\affiliation{Department of Physics and Astronomy and National Superconducting Cyclotron Laboratory, Michigan State University, East Lansing, Michigan, 48824-1321, USA}

\author{T.~Redpath}
\affiliation{Department of Physics, Central Michigan University, Mt. Pleasant, Michigan 48859, USA}

\author{T. Roger}
\affiliation{Grand Acc\'el\'erateur National d'Ions Lourds (GANIL),
CEA/DSM - CNRS/IN2P3, B.\ P.\ 55027, F-14076 Caen Cedex 5, France}

\author{F.~ Rotaru}
\affiliation {IFIN-HH, P. O. Box MG-6, 76900
Bucharest-Magurele, Romania}

\author{S.~Saenz}
\affiliation{Department of Physics, Central Michigan University, Mt. Pleasant, Michigan 48859, USA}

\author{M.-G.~Saint-Laurent}
\affiliation{Grand Acc\'el\'erateur National d'Ions Lourds (GANIL),
CEA/DSM - CNRS/IN2P3, B.\ P.\ 55027, F-14076 Caen Cedex 5, France}

\author{D.~Smalley}
\affiliation{National Superconducting Cyclotron Laboratory, Michigan State University, East Lansing, Michigan 48824, USA}

\author{D.\ Sohler}
\affiliation{Institute of Nuclear Research of the
Hungarian Academy for Sciences, P.O. Box 51, Debrecen, H-4001, Hungary}

\author{M.\ Stanoiu}
\affiliation {IFIN-HH, P. O. Box MG-6, 76900
Bucharest-Magurele, Romania}

\author{S.~R.~Stroberg}
\affiliation{Department of Physics and Astronomy and National Superconducting Cyclotron Laboratory, Michigan State University, East Lansing, Michigan, 48824-1321, USA}

\author{J.C.~Thomas}
\affiliation{Grand Acc\'el\'erateur National d'Ions Lourds (GANIL),
CEA/DSM - CNRS/IN2P3, B.\ P.\ 55027, F-14076 Caen Cedex 5, France}

\author{M. Vandebrouck}
\affiliation{Institut de Physique Nucl\'eaire, IN2P3-CNRS,
F-91406 Orsay Cedex, France}

\author{D. Weisshaar}
\affiliation{Department of Physics and Astronomy and National Superconducting Cyclotron Laboratory, Michigan State University, East Lansing, Michigan, 48824-1321, USA}

\author{A.~Westerberg}
\affiliation{Department of Physics, Central Michigan University, Mt. Pleasant, Michigan 48859, USA}

\begin{abstract}
Excited states in $^{28}$Na have been studied using the $\beta$-decay of implanted $^{28}$Ne ions at GANIL/LISE as well as  the in-beam $\gamma$-ray spectroscopy at the NSCL/S800 facility. New states of positive (J$^{\pi}$=3,4$^+$) and negative (J$^{\pi}$=1-5$^-$) parity are proposed. The former arise from the coupling between 0d$_{5/2}$ protons and a 0d$_{3/2}$ neutron, while the latter are due to couplings with 1p$_{3/2}$ or 0f$_{7/2}$ neutrons. While the relative energies between the  J$^{\pi}$=1-4$^+$ states are well reproduced with the USDA interaction in the N=17 isotones, a progressive shift in the ground state binding energy (by about 500 keV) is  observed between $^{26}$F and $^{30}$Al. This points to a possible change in the proton-neutron 0d$_{5/2}$-0d$_{3/2}$ effective interaction when  moving from stability to the drip line.  The presence of J$^{\pi}$=1-4$^-$ negative parity states  around 1.5 MeV as well as of a candidate for a J$^{\pi}$=5$^-$ state around 2.5 MeV give further support to the collapse of the N=20 gap and to the inversion between the 0f$_{7/2}$ and 1p$_{3/2}$ levels below Z=12. These features are discussed in the framework of Shell Model and EDF calculations, leading to predicted negative parity states in the low energy spectra of the $^{26}$F and  $^{25}$O nuclei.

\end{abstract}

\pacs{21.60.Cs, 23.20.Lv, 27.30.+t} 
\maketitle

\section{Introduction}
The first disappearance of a magic shell was proposed more than 30 years ago for the neutron magic number N=20. This discovery arose from the  combined works on atomic masses~\cite{Thib75}, nuclear radii~\cite{Hube78} and nuclear spectra ~\cite{Detr79,Guil84} of nuclei around $^{32}$Mg. This discovery was later confirmed by complementary measurements on the reduced transition probability values B(E2;0$^+ \rightarrow$ 2$^+$) \cite{N20BE2}, on the  quadrupole and magnetic moments \cite{N20gfact},  as well as on the neutron knock-out cross sections \cite{N20knock}, to quote a few.  Theoretically  the works of Refs~\cite{Camp75,Pove87,Warb90} described this onset of collectivity at N=20 as due to the combination of a shell gap reduction and the excitations of particles from the normally occupied orbital to the first orbital of the upper shell. These excitations lead to a significant increase of correlations, eventually bringing the \emph{intruder} configuration from the upper shell below the normal configuration.  The nuclei for which the ordering of the intruder and normal configurations is inverted belong to the so-called 'Island of Inversion'~\cite{Warb90}. Their configurations are often strongly mixed.

\begin{figure}
\centering \epsfig{width=6cm,file=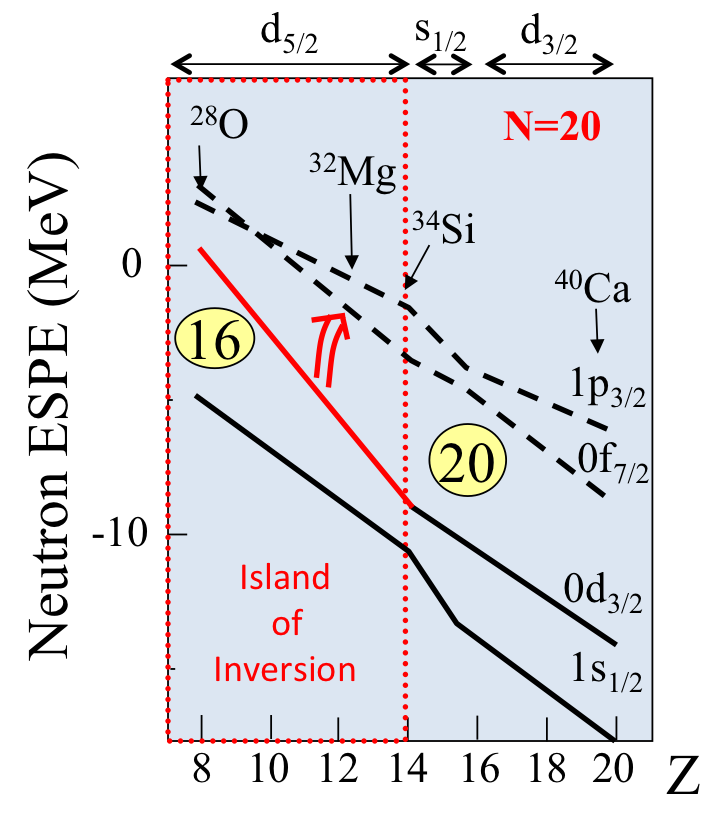}
\caption{(color on line) Effective neutron single-particle energies (ESPE) of the N=20 isotones between Z=8 and Z=20 based on the work of Ref.\cite{Utsu}. The slope of the lines corresponds to the strength of the proton-neutron monopole interactions. The red line, which has the largest slope, shows the effect of the strong d$_{5/2}$-d$_{3/2}$ proton-neutron monopole interaction when the proton d$_{5/2}$ is filled.}
\label{ESPEN}
\end{figure}

The shell evolution of neutron orbits as a function of the proton number is illustrated in Fig. \ref{ESPEN} for the N=20 isotones. This figure has been elaborated with the monopole proton-neutron interactions obtained in the shell model approach of Ref. \cite{Utsu} that include the few experimental observations made at that time. At Z=8, the neutron 0d$_{3/2}$, 1p$_{3/2}$ and 0f$_{7/2}$ orbits are unbound, the N=20 is weaker than the N=16 gap, and the ordering of the 1p$_{3/2}$ and 0f$_{7/2}$ orbits is reversed compared to what is observed in the valley of stability. While filling the proton 0d$_{5/2}$ orbit from Z=8 ($^{28}$O) to Z=14 ($^{34}$Si), the N=20 gap increases as the N=16 gap is reduced. After Z=14, the filling of the 1s$_{1/2}$ and 0d$_{3/2}$ proton orbits  keeps the N=20 gap unchanged between Z=14 and Z=20. The formation of the N=20 shell gap, as well as the inversion between the neutron  0f$_{7/2}$ and 1p$_{3/2}$ orbitals  are profound structural changes in the shells that are likely caused by the hierarchy between the proton-neutron monopole interactions involved: the 0d$_{5/2}$-0d$_{3/2}$ interaction is much larger than the others involved  such as the 0d$_{5/2}$-0f$_{7/2}$ and 0d$_{5/2}$-1p$_{3/2}$ ones, as described in Refs. \cite{hierarchy,Utsu,Otsu05}. When the 1p$_{3/2}$ orbit becomes unbound a further reduction of the monopole interactions involving this low-$\ell$ orbital may be present.

The 0d$_{5/2}$-0d$_{3/2}$, 0d$_{5/2}$-1p$_{3/2}$ and 0d$_{5/2}$-0f$_{7/2}$  monopole interactions are so far  poorly constrained by experimental data. To achieve this goal, the evolution of the 0d$_{3/2}$, 1p$_{3/2}$ and 0f$_{7/2}$ neutron single-particle  energies should be determined from the Si to O isotones.  However the N=20 nuclei lying in the island of inversion are deformed. Consequently their configurations are strongly mixed and their spherical single-particle energies can rarely be determined. A further difficulty arises from the fact that the N=20 nuclei with Z=10 and Z=8 are weakly bound or unbound. They are difficult to produce in experiments and their study would often require the treatment of the continuum. In order to study these interactions and their behavior toward the neutron drip line, we take advantage of the fact that large N=14 and N=16 shell gaps are present around Z=8-10 \cite{Thir00,Stan04,Oza00,Bech,Hoff,Hoff08,Tshoo}, where the N=20 gap is weakened. The N=17 nuclei can therefore be described mainly with a single neutron in the d$_{3/2}$, p$_{3/2}$ or f$_{7/2}$ orbit, without a large mixing with neighboring shells.

Recent experimental observations \cite{Catf,ne27,Terry06,Ober06} have brought remarkable credit to the theoretical description of Fig.\ref{ESPEN} around (N=16, Z=8) suggesting an inversion between the 1p$_{3/2}$ and 0f$_{7/2}$ orbits in the N=15 and N=17 Ne isotopes. These experiments have also demonstrated that the N=20 gap is small. In addition, the works of \cite{Hoff,Hoff08,Tshoo} have proven that the N=16 shell gap is large at Z=8, while Ref. \cite{Lepailleur} suggests that the N=16 gap somehow persists at Z=10 in $^{26}$Ne, as witnessed by its vibrational behavior.

The  best way to extract information about the aforementioned monopole \textit{and} multipole parts of the nuclear interaction is therefore  to study odd-odd nuclei in the N=17 isotones. The coupling of protons and neutrons in the d$_{5/2}$ and d$_{3/2}$ orbits will lead to J$^{\pi}$=1-4$^+$ states in the $^{30}_{13}$Al$_{17}$, $^{28}_{11}$Na$_{17}$ and $^{26}_{9}$F$_{17}$ isotones that  span from near stability to the neutron drip line. In a similar manner, the coupling of the protons in the d$_{5/2}$ orbit with neutrons in the  1p$_{3/2}$ (0f$_{7/2}$) orbit leads to J$^{\pi}$=1-4$^-$ (J$^{\pi}$=1-6$^-$) negative parity states  in the same nuclei. However such studies are tedious as several experimental methods are often required to produce all the states of these multiplets for nuclei that are not so easily produced at radioactive ion beam facilities. The first J=1-4 positive parity states in $^{30}$Al have been obtained recently using the Gammasphere array \cite{Step}. Negative parity states are proposed from 2.29 MeV on, but they were  assumed to originate from the neutron f$_{7/2}$ orbital. In $^{26}$F three different experimental techniques were required to study the bound J$^{\pi}$=2,4$^+$ states \cite{Sta12,Lepailleur} and the unbound J$^{\pi}$=3$^+$ state \cite{Fran11}. So far there is no evidence of negative parity states in $^{26}$F below the neutron emission threshold of 1.070(62) MeV \cite{Lepailleur}.  As for the $^{28}$Na nucleus, there are only candidates for the $J^{\pi}=1_1^+$ and $J^{\pi}=2_1^+$ states that were derived from a previous $\beta$-decay study of $^{28}$Ne \cite{Tripa}. The present study aims at determining the energy of the missing $J^{\pi}=3_1^+$ and $J^{\pi}=4_1^+$ states and providing information on the presence of negative parity states in $^{28}$Na to confirm the lowering of the 1p$_{3/2}$ and 0f$_{7/2}$ orbits toward Z=8. To achieve these goals two complementary experimental techniques were required. We first repeated the $\beta$-decay experiment of  $^{28}$Ne at the GANIL/LISE facility with a larger statistics as compared to \cite{Tripa} and we secondly used the in-beam spectroscopy technique at the NSCL/MSU facility to detect the $\gamma$-rays of $^{28}$Na in the GRETINA  Ge detector array \cite{Gretina} produced in the neutron and proton removal reactions from $^{31,32}$Mg.

\section{$\beta$-decay experiment }
\subsection{Experimental technique}

\begin{figure}
\centering \epsfig{width=8.5cm,file=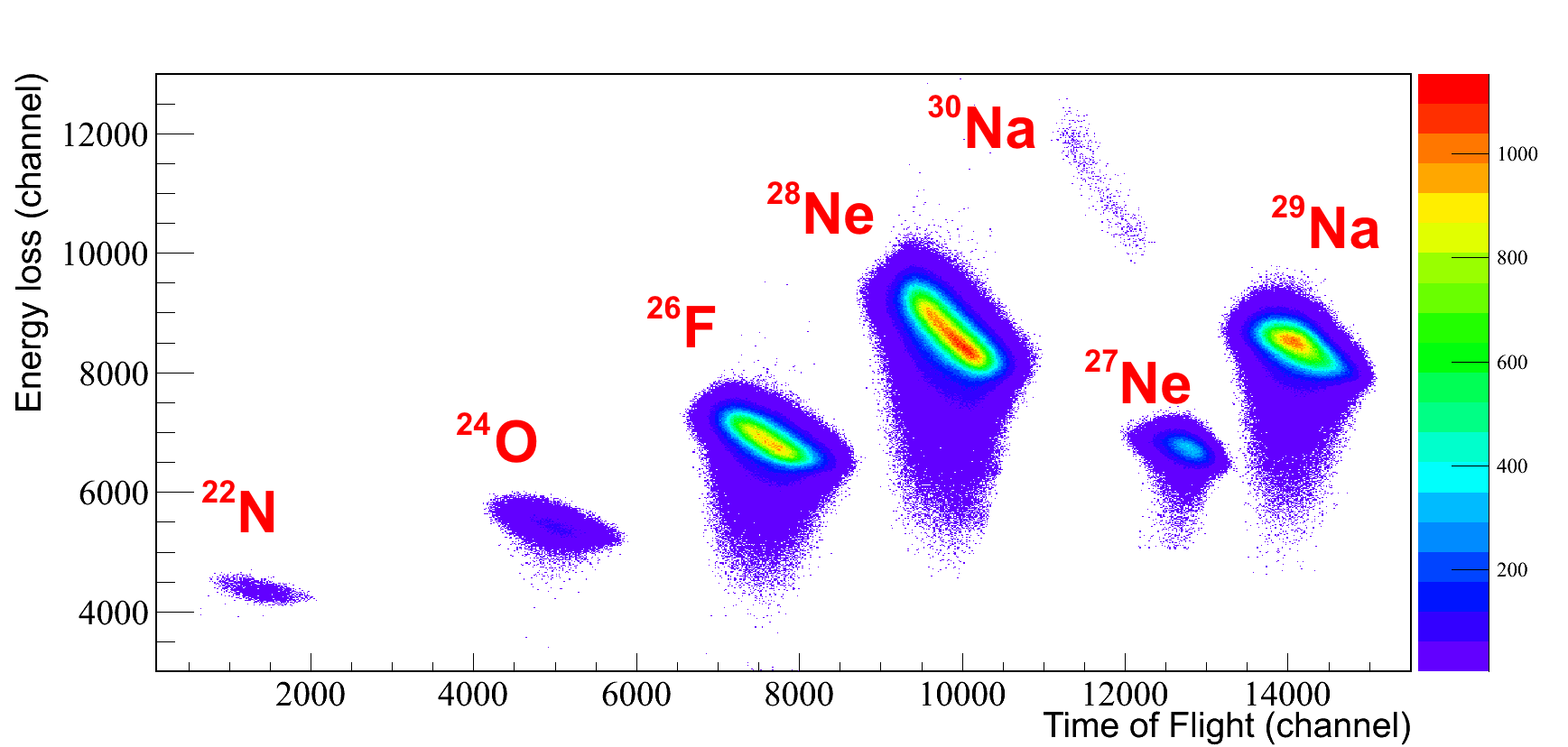}
\caption{ Identification of the nuclei produced in the experiment through their energy loss and time of flight.}
\label{ions_id}
\end{figure}

The $^{28}$Ne nuclei were produced at the GANIL facility through the fragmentation of a 77.6 $\it{A} \cdot$ MeV $^{36}$S$^{16+}$  beam with an intensity of 2 $\mu$Ae
on a 237 mg/cm$^2$ Be target. Nuclei of interest were selected by the LISE \cite{LISE}
spectrometer, in which a wedge-shaped Be degrader of 1066 $\mu$m
was inserted at the intermediate focal plane. They were identified from their energy losses in a stack of three 500 $\mu$m Si
detectors and from their time-of-flight referenced to the cyclotron radio frequency. The spectrometer set-up was  optimized for $^{26}$F, but  a fraction of the $^{28}$Ne ions were also transmitted at a rate of 10~pps, corresponding to 22\% of the implanted nuclei shown in Fig. \ref{ions_id}. A total of 3.24 $\times$10$^6$   $^{28}$Ne  nuclei were implanted in a 1~mm-thick double-sided Si stripped detector (DSSSD) composed of 256 pixels (16 strips in both the X and Y directions) of 3$\times$3~mm$^2$  located at the
final focal point of LISE.  $\beta$-particles were detected in the same strip of the DSSSD as the precursor nucleus $^{28}$Ne.

\begin{figure*}
\centering \epsfig{width=18cm,file=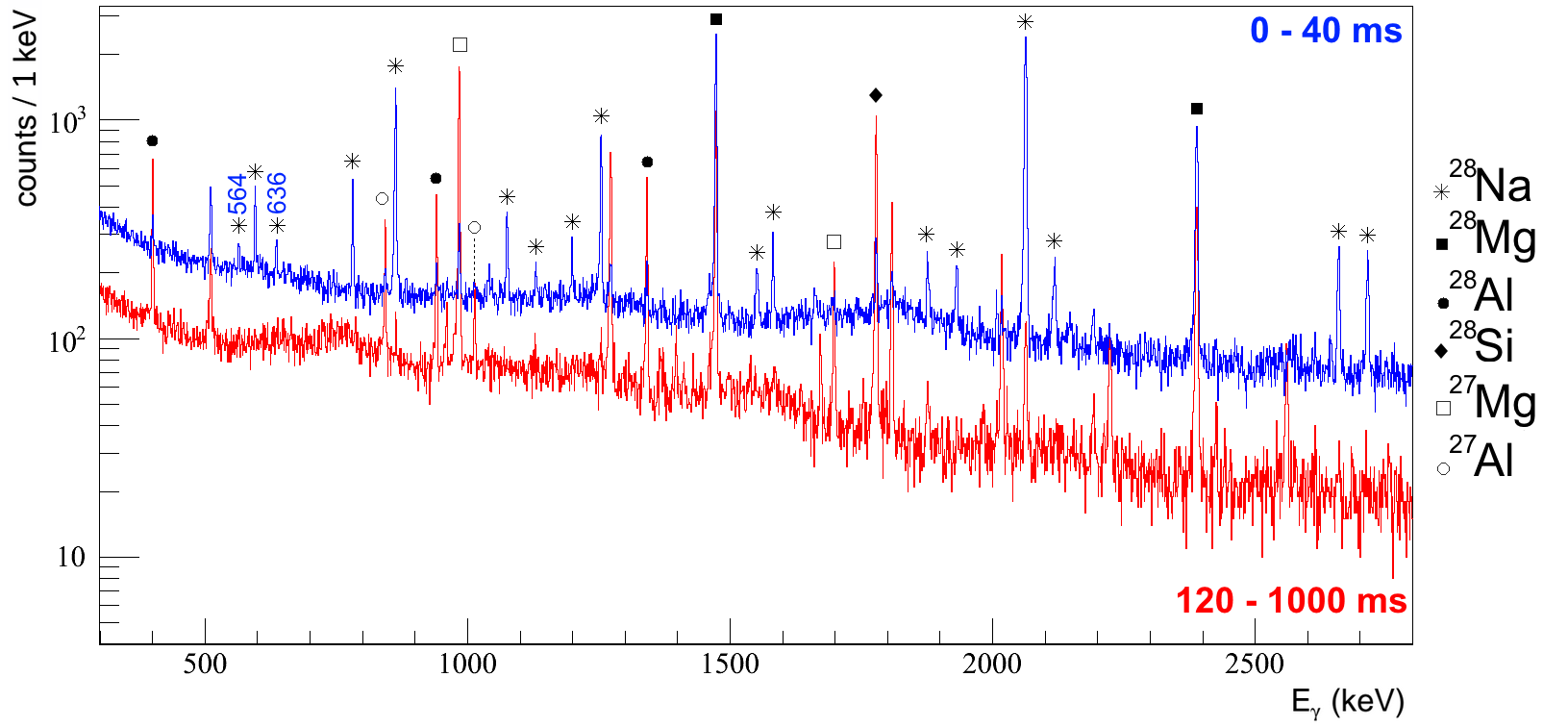}
\caption{(color on line): $\beta$-gated $\gamma$-ray spectra obtained up to 40 ms (blue, upper) and between 120 ms and 1000 ms (red, bottom) after the implantation of a  $^{28}$Ne nucleus. The red histogram has been scaled down to see the two spectra on top of each others. Identified $\gamma$-rays are shown with different symbols. The new ones at 564 (1) and 636 (1) keV attributed to the decay of $^{28}$Ne are visible on the left hand side of the spectrum. }
\label{Ege_beta_tot}
\end{figure*}
Four Ge detectors of the EXOGAM  array \cite{EXOGAM} surrounded the DSSSD to detect
the $\gamma$-rays with an efficiency of 6.5(3)\% at 1~MeV. A total $\beta$-efficiency of 60(1)\% has been determined
from the comparison of the intensity of a given $\gamma$-ray belonging to the decay of $^{28}$Ne gated or not on a $\beta$-ray.
A Si-Li detector of 5 mm thickness was placed downstream of the DSSSD for three purposes: i)  to ensure that the $^{28}$Ne ions were implanted in the DSSSD and not passing through it, ii) to discriminate  very light particles in the beam that pass through the telescopes and iii) to determine the $\beta$ energy threshold in each strip using the coincidence with the $\beta$-particles detected in Si-Li detector.

\begin{figure}
\centering \epsfig{width=8.5cm,file=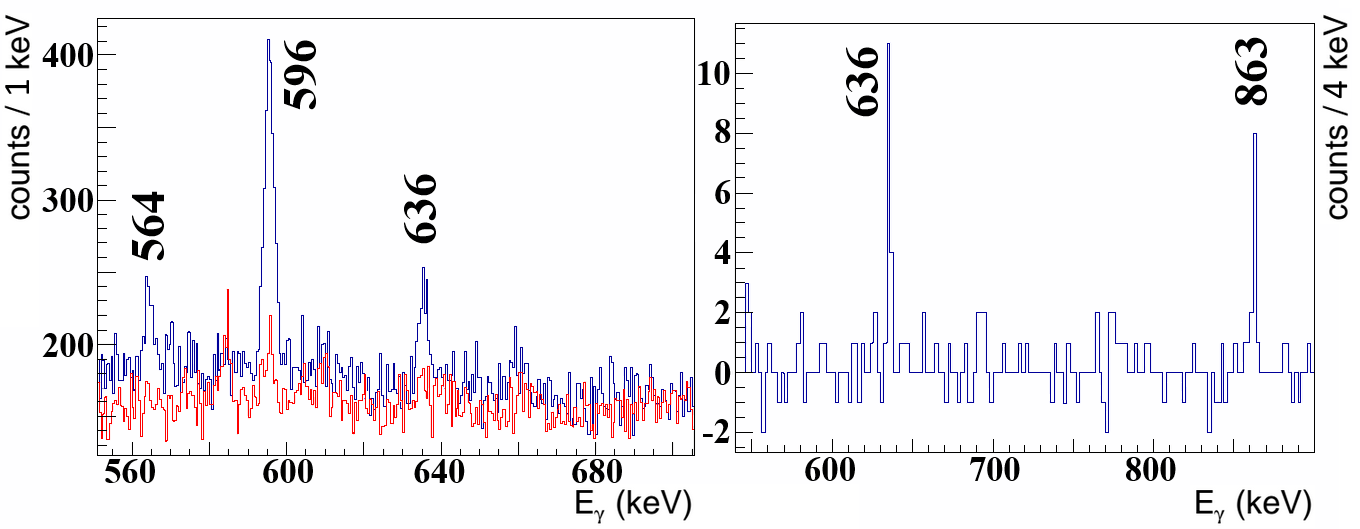}
\caption{(color on line) Left: $\beta$-gated $\gamma$-ray spectrum obtained up to 60 ms (blue) and between 120 ms and 1000 ms (red) after the implantation of a  $^{28}$Ne nucleus. Right: $\beta$-gated $\gamma$-ray spectrum in coincidence with the 564 (1) keV transition.}
\label{Ege_beta_zoom+coincs}
\end{figure}

\subsection{Beta-decay scheme of $^{28}$Ne}
The upper (bottom) spectrum of Fig. \ref{Ege_beta_tot} displays the $\gamma$-ray energy spectrum obtained with a $\beta$-particle correlated in space and time up to 40 ms (between 120 and 1000 ms) after the implantation of a $^{28}$Ne precursor. Owing to the short lifetime of  $^{28}$Ne (see below), the transitions from Fig. \ref{Ege_beta_tot} belonging to its $\beta$-decay are more intense in the upper (blue) spectrum, while those caused by daughter decays or other implanted nuclei are dominating the bottom (red) spectrum.   Consistent half-lives values of
T$_{1/2}$=18.2(5) ms and 18.6 (2) ms are found from the $\beta$-decay time spectrum of $^{28}$Ne gated on the 863 keV and 2063 transitions seen in Fig. \ref{Ege_beta_tot}, respectively. These values are also in accordance with T$_{1/2}$($^{28}$Ne) = 18.4 (5) ms derived in Ref. \cite{Tripa}.

The level scheme shown in the left part of Fig. \ref{schema_lvl} is established from $\beta$-gated $\gamma$-$\gamma$ coincidences following the implantation of a $^{28}$Ne nucleus. Based on their direct $\beta$-decay feeding from the 0$^+$ ground state of $^{28}$Ne and Gamow-Teller $\beta$-decay selection rules, we propose a J$^{\pi}$=1$^+$ spin parity value to the states at 2714, 2118 keV and to the ground state of $^{28}$Na.  The $\beta$ feedings to the $^{28}$Na ground state as well as to unbound states (leading to the $\beta$-delayed neutron emission) are derived from the intensities of the observed lines seen in Fig. \ref{Ege_beta_tot} populating excited states in the $^{28}$Mg and $^{27}$Mg nuclei, respectively. They agree with the experimental feedings determined in Ref. \cite{Tripa}  for most of the states, including for unbound states. As shown in Fig.\ref{schema_lvl}, these $\beta$ feedings compare reasonably well with shell model calculations. However, as for the $\beta$-decay of $^{26}$Ne \cite{Weis04} the feeding of only three of the four predicted 1$^+$ states  is observed in the $\beta$-decay of $^{28}$Ne below the neutron separation energy.

\begin{table}
\caption{List of states, $\gamma$-ray energies and relative intensities (normalized to 100 $^{28}$Ne decays) observed in the $\beta$-decay of  $^{28}$Ne to $^{28}$Na.}
\begin{ruledtabular}
\begin{tabular}{rrrr}
$E_\text{i}$ (keV) & $E_\text{f}$ (keV) & $E_\gamma$ (keV) & $I^\text{rel}$ (\%) \\
\hline
  55 \footnotemark[1]\footnotetext[1]{taken from Ref. \cite{Tripa}} &    0 &   55 & - \\
 691(1) &   55 &  636(1) & 0.3(1) \\
 1131(1) &   0 &  1131(1) &  0.2(1) \\
     &  55 &  1076(1) &  1.0(1) \\
1255(1) &    0 & 1255(1) &   2.7(2) \\
     &   55 & 1200(1) & 0.5(1) \\
     &  691 &  564(1) &   0.3(1) \\
1932(1) &    0 & 1932(1) &  0.5(1) \\
       & 55 &  1877(2) &   0.4(1) \\
2118(1) &   0 & 2118(2) &  0.8(1) \\
 & 55 &  2063(1) &   14.2(12) \\
 & 55 &  863(1) &   3.6(3) \\
2714(1) &   0 & 2714(2) &  0.9(1) \\
 & 55 &  2659(1) &   0.9(1) \\
 & 1131 &  1583(1) &   0.7(1) \\
 & 1932 &  782(1) &   1.1(1) \\
 & 2118 &  596(1) &   0.6(1) \\
\end{tabular}
\label{Tab1}
\end{ruledtabular}
\end{table}

With the exception of the 3231 keV and 3457 keV transitions, all the transitions belonging to the $^{28}$Ne decay to $^{28}$Na observed in \cite{Tripa} are observed here. With $\beta$ feedings of $\simeq$ 2 \%  for the 3231 keV and 3457 keV transitions given in Ref.  \cite{Tripa}, they should have been seen in our data with a confidence level of 10 $\sigma$. As the spatial correlation was not used in Ref. \cite{Tripa}, it is possible  that these $\gamma$ rays were wrongly assigned to the $^{28}$Ne decay.  We observe two new $\gamma$ rays at 564(1) keV and 636(1) keV of similar weak intensity that are in mutual coincidence (see Fig. \ref{Ege_beta_zoom+coincs}) as well as in coincidence with the 863(1) keV $\gamma$ rays de-exciting  the 2118(1) keV level in $^{28}$Na. The summed  energy of these two $\gamma$ transitions, 1200(2) keV, matches the energy of the 1200(1) keV transition coming from the decay of the 1255(1) keV to 55.2(5) keV states in $^{28}$Na. We therefore propose a new level in $^{28}$Na, the placement of which  (691(2) keV) was derived using the information obtained in a second experiment described below.  The list of states, $\gamma$-ray energies and intensities observed in the $\beta$-decay experiment is given in Table ~\ref{Tab1}.

To propose spin parity assignments to the identified states, we start with the fact the ground state of $^{28}$Na has J$^{\pi}$=1$^+$, which is deduced from its large direct feeding from the 0$^+$ ground state of $^{28}$Ne. The excited state at 55 keV is likely to have J$^{\pi}$=0,2$^+$ as a larger spin difference between this 55 keV state and the ground state would have resulted in a long-lived isomer transition neither observed in the $\beta$-decay part of our work nor in Ref.~\cite{Tripa}.  Between the J$^{\pi}$=0,2$^+$ candidates, the J$^{\pi}$=2$^+$ value seems the most reasonable, considering that the USDA \cite{usdab} prediction gives the first 2$^+$ state around 100 keV and the first 0$^+$ state at a much higher energy of about 2 MeV. Based on the comparison to shell model calculations and their respective feeding from higher energy states and decay branches to the 55 keV or ground state, the state at 691 keV is a good candidate for J$^{\pi}$=3$^+$. Indeed a J$^{\pi}$=3$^+$ state mainly decays through an M1 transition to the J$^{\pi}$=2$^+$ at 55 keV with a $\gamma$-ray of 636 keV rather than through an E2 transition to the J$^{\pi}$=1$^+$ ground state. We propose spin parity assignments J$^{\pi}$=3$^+_2$ and J$^{\pi}$=2$^+_2$ to the states at 1131 keV and 1255 keV from the comparison to shell model calculations as well as from their feeding and decay branching ratios in E2 and M1 transitions. The 1932 keV state likely  has a J$^\pi$=2$^+$  configuration as it decays equally to the 1$^+$ ground state and to the 2$^+$ state at 55 keV. Most of the states predicted by the shell model calculations using the USDA interaction have their experimental equivalent up to 1255 keV.

\section{In-beam  $\gamma$-ray  spectroscopy  }
\subsection{Experimental technique}
\begin{figure}
\centering\includegraphics[width=\columnwidth]{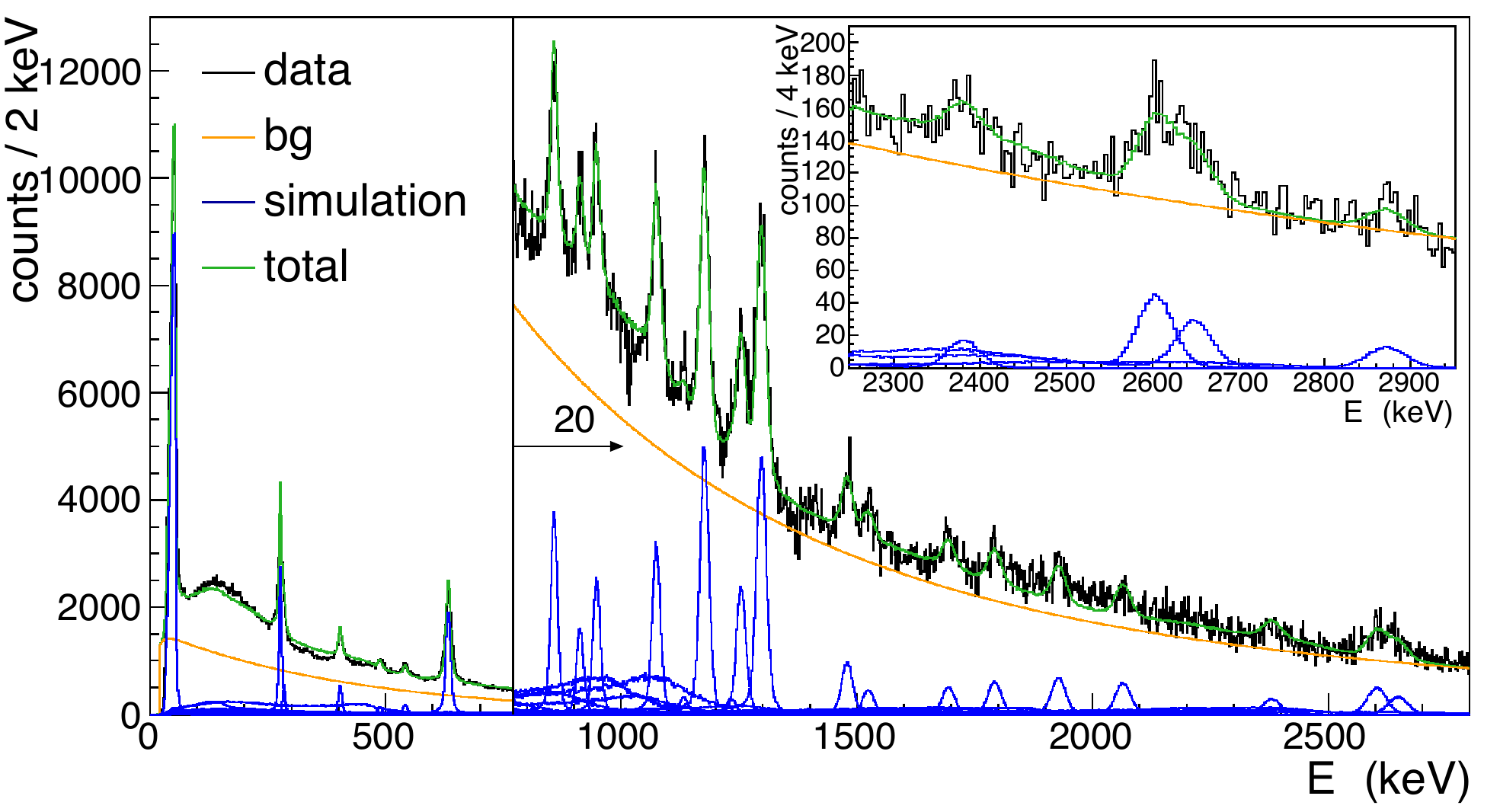}
\caption{(color on line) Doppler-reconstructed $\gamma$-ray energy spectrum obtained from in-beam $\gamma$-decay spectroscopy of $^{28}$Na. The solid green line is the response obtained from the GEANT4 simulation if individual transitions (blue lines) and a continuous background (orange line) are considered. The inset shows the high energy part of the spectrum with in particular the doublet of transitions at 2605 and 2650 keV.}
\label{Ege_MSU}
\end{figure}

In-beam $\gamma$-ray spectroscopy of the neutron-rich $^{28}$Na isotope was performed in a second experiment at the Coupled Cyclotron Facility of NSCL at Michigan State University. The $^{28}$Na nuclei were produced in the secondary fragmentation reactions from $^{31}$Mg and $^{32}$Mg beams impinging at about 95 MeV/u on a 375~mg/cm$^2$ $^9$Be target. In total about $7.7\cdot10^5$ $\gamma - ^{28}$Na coincidences were recorded in the two settings together. The  outgoing particles were identified  based  on  the   time-of-flight   and  energy-loss
measurements  using  the  focal-plane  detection system  of  the  S800
spectrograph~\cite{S800}.  Trajectories of recoiling ions were tracked
in the S800  using two sets of Cathode  Readout Drift Chambers measuring their position  and angle values  at the focal plane allowing
the  reconstruction of  their velocity  ($\beta \simeq$ 0.41) and position at the  target.  The target   was   surrounded   by    seven   modules   of   the   GRETINA
array~\cite{Gretina}, each module consisting in four HPGe segmented
crystals.  Four modules were placed  at the most forward angles around
58$^\circ$ and  three around 90$^\circ$  to  detect $\gamma$ rays
induced by the de-excitation of the nuclei in-flight with an efficiency
of about  6 \% at  1 MeV. The $\gamma$-ray detection threshold has been lowered to about 50 keV in order to be able to observe the low-lying transition from the first excited state in $^{28}$Na at 55~keV. The threshold and energy resolutions were modeled using the GEANT4~\cite{agostinelli03} simulation of the GRETINA setup~\cite{riley14}. The $^{28}$Na  velocity vector and position were  used to
apply an event-by-event Doppler correction to the $\gamma$-ray spectrum
of the  $^{28}$Na nuclei  shown in Fig. \ref{Ege_MSU}.   A $\gamma$-ray
energy resolution  (FWHM) of about  2\% at  1 MeV was  achieved. The list of $\gamma$-ray energies and intensities observed in this experiment is given in Table \ref{Tab2}.

\begin{table}[h]
\caption{List of states, $\gamma$-ray energies and intensities observed in the in-beam experiment.
Systematic uncertainties $\sigma(E)$ of about 3 keV should be considered on the proposed $\gamma$-ray energies.}
\begin{ruledtabular}
\begin{tabular}{rrrr}
$E_\text{i}$ (keV) & $E_\text{f}$ (keV) & $E_\gamma$ (keV) & $I^\text{rel}$ (\%)\footnote{normalized to the 633~keV transition} \\
\hline
  55 &    0 &   55 & 380(50) \\
 688 &   55 &  633 & 100(2) \\
 968 &   55 &  912 &   7(1) \\
     &  688 &  277 &  52(3) \\
1131 &    0 & 1131 &   1.9(6) \\
     &   55 & 1075 &  16.6(8) \\
1233 &    0 & 1233 &   1.9(6) \\
     &   55 & 1177 &  29(1) \\
     &  688 &  542 &   8.3(6) \\
1255 &    0 & 1255 &  15.7(8) \\
1353 &   55 & 1298 &  33(1) \\
1481 &    0 & 1481 &   8.2(8) \footnotemark[2]\footnotetext[2]{placement uncertain}\\
1636 &  688 &  948 &  11.6(8) \\
     & 1233 &  403 &  16.7(7) \\
     & 1353 &  282 &  10(2) \\
1749 &   55 & 1694 &   5.2(7) \\
1792 &    0 & 1792 &   6.9(9) \\
1929 &    0 & 1929 &   8.5(9) \footnotemark[3]\footnotetext[3]{placement based on $\beta$ decay} \\
2121 &   55 & 2066 &   8.2(8) \footnotemark[3] \\
2378 &    0 & 2378 &   5.2(8) \footnotemark[4] \\
2493 &  965 & 1527 &   4(1) \\
     & 1636 &  858 &  14.8(8) \\
2605 &     & 2605 &  10(1) \footnotemark[4]\\
2650 &     & 2650 &   7(1) \footnotemark[4]\\
2874 &     & 2874 &   4(1) \footnotemark[4]\\
     &      &  489 &  5.7(6) \footnotemark[4]\footnotetext[4]{unplaced}\\
\end{tabular}
\label{Tab2}
\end{ruledtabular}
\end{table}
The systematic uncertainty for the intensity of the 55 keV transition is particularly large because of the vicinity of the threshold. Additionally, this state exhibits a relatively long lifetime as deduced from its low energy tail. It follows that the Doppler shifted energy measured in GRETINA lies partially below the detection threshold. The lifetime of the $2^+$ state has been simulated, and the resulting response function has been fitted to the experimental spectrum. This way a lower limit for the lifetime of $\tau > 1$~ns is obtained. An additional constraint on the lifetime can be obtained from the fact that 100~\% of the 633~keV $\gamma$-ray yield has to proceed through the 55~keV transition. Consistent yields in the 633-55 keV coincidence are obtained for simulated lifetimes of $\tau = 1.4(4)$~ns.
Shell model calculations using the USDA effective interaction predict the $2^+$ state at 182~keV. Using the shell model reduced transition probabilities, B(E2) and B(M1) values, for the decay to the ground state, and the experimental transition energy of 55~keV, a theoretical lifetime of $\tau_\text{theo} = 1.42$~ns is obtained. This validates our assumption of a relatively long lifetime of the $2^+$ state at 55~keV.

\begin{figure}
\centering\includegraphics[width=\columnwidth]{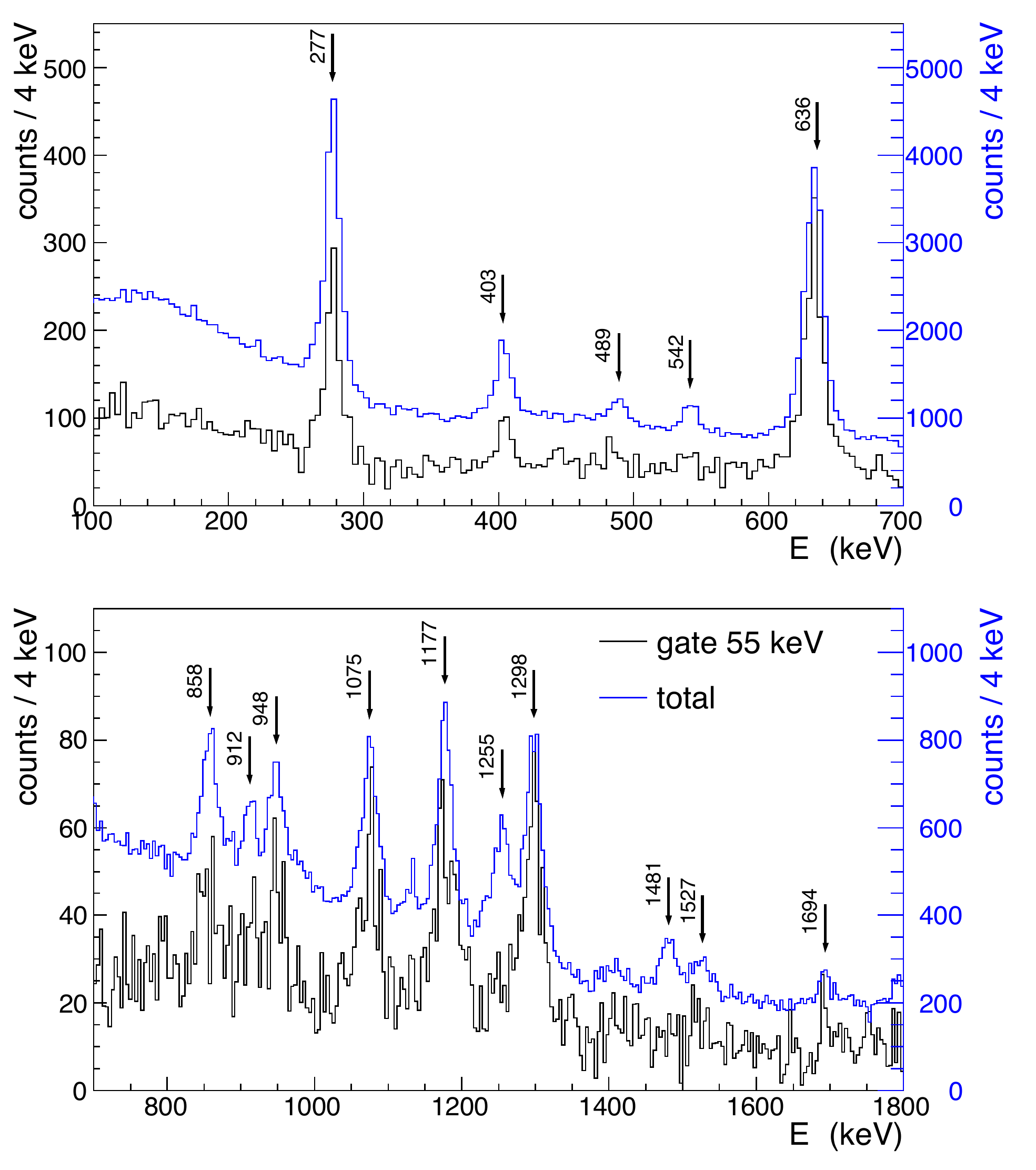}
\caption{(color on line) Doppler-reconstructed $\gamma$-ray energy spectrum gated on the 55~keV transition from the first excited state in $^{28}$Na in two different energy ranges. The blue line (top spectrum in each pannel) shows the total spectrum for comparison. While most of the transitions are observed in coincidence with the 55~keV $\gamma$ ray, the transitions of 1255 and 1481~keV are for instance clearly missing in the $\gamma$-gated spectrum.}
\label{Ege_MSU_coinc}
\end{figure}

Due to the high detection efficiency of the GRETINA array it was possible to construct $\gamma - \gamma$ coincidence spectra by gating on several transitions. Fig.~\ref{Ege_MSU_coinc} shows the spectrum observed in coincidence with the 55~keV transition.
This spectrum  shows which of the  observed  $\gamma$ transitions are directly or indirectly populating the first excited state at 55~keV.

\begin{figure}
\centering \epsfig{width=8.5cm,file=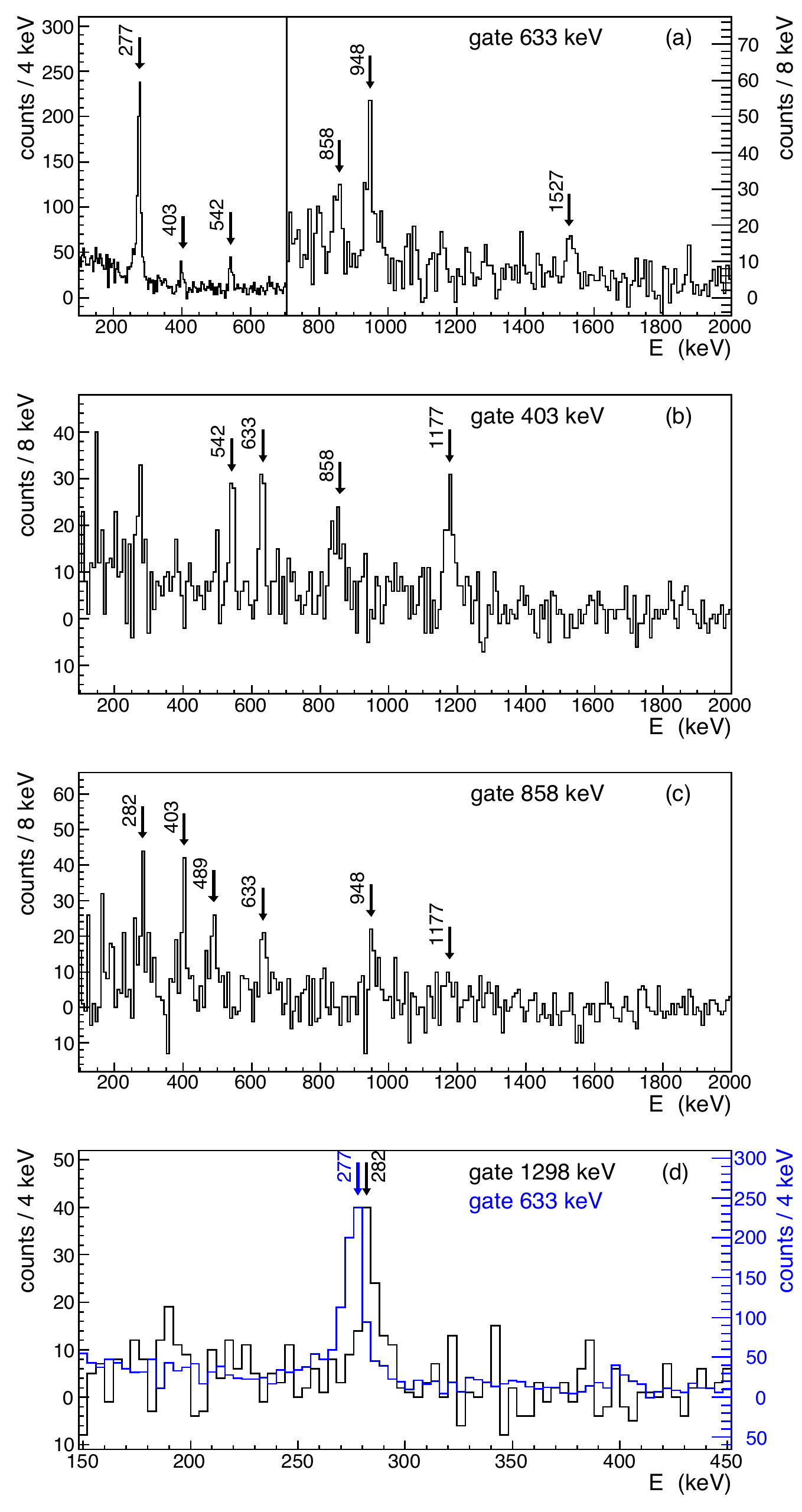}
\caption{ (color on line) $\gamma-\gamma$ coincidence spectra of $^{28}$Na obtained during the in-beam experiment when gating on several $\gamma$-rays (from top to bottom 633, 403, 858 and (633,1298) keV).}
\label{Ege_MSU_coinc2}
\end{figure}

\subsection{Level scheme of $^{28}$Na }
\begin{figure*}
\centering \epsfig{width=18cm,file=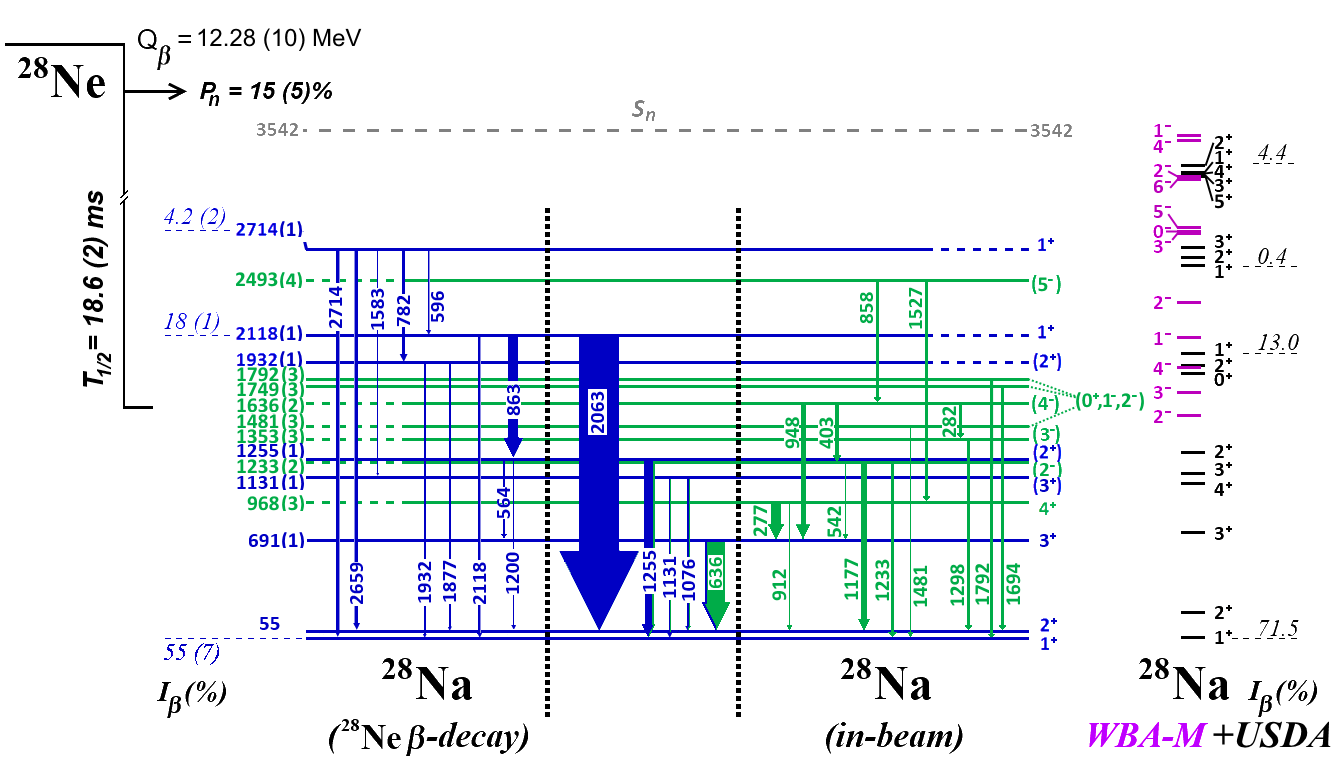}
\caption{(color on line) The very right part displays shell model predictions using the USDA (WBA-M) interaction for the positive (negative) parity states. The rest shows the experimental level scheme of  $^{28}$Na derived by adding information from the $\beta$-decay of $^{28}$Ne (left) and from the in-beam gamma-ray spectroscopy (right). The middle part displays $\gamma$-lines that are observed in the two experiments. The energies and uncertainties of the levels are derived from the $\beta$-decay when possible (Table \ref{Tab1}), from the in-beam experiment (Table \ref{Tab2}), or from a combination of the two experiments otherwise.}
\label{schema_lvl}
\end{figure*}

The level scheme of $^{28}$Na as well as tentative spin assignments derived from this part of the experiment are based on the single $\gamma$-ray energy spectrum of Fig. \ref{Ege_MSU},  $\gamma$-$\gamma$ coincidence spectra of Fig.~\ref{Ege_MSU_coinc} and \ref{Ege_MSU_coinc2} as well as $\gamma$ branching ratios from Table~\ref{Tab2}.  Two almost overlapping $\gamma$ transitions are present at  277 and 282 keV. The 633(3) keV  $\gamma$-ray  is compatible  with  the  636(1) keV $\gamma$-ray  observed  in   the  $\beta$-decay  experiment.  As  the 564(1) keV $\gamma$-ray found in the $\beta$-decay experiment is not observed here, it is
placed above the 636(1) keV transition.  The 277(1) keV  $\gamma$-transition is  in
coincidence with  the 636 keV one. As the intensity of the latter is
larger, it is placed below  the 277 keV transition, establishing a new state  at 968(3)  keV as seen in Fig.~\ref{schema_lvl}.  A tentative J$^{\pi}$=4$^+$ assignment is proposed for this level as it mainly decays by a 277 keV $\gamma$-ray to the previously assigned 3$^+$ at 691 keV, very weakly to the 55 keV 2$^+$ state and not to the ground state with J$^{\pi}$=1$^+$. With this newly proposed J$^{\pi}$=4$^+$ state, all positive parity states but the J$^{\pi}$=0$^+$ that are predicted by the shell model calculations below 2.2 MeV are observed experimentally. As candidates for J$^{\pi}$=0$^+$ state are proposed in the following paragraph most of the other states populated in this reaction and shown in the right part of Fig.~\ref{schema_lvl} are proposed to be candidates for intruder states, arising from the neutron $fp$ shells.

A tentative J$^{\pi}$=2$^-$ assignment is proposed to the 1233 keV level from the fact that it decays to the J$^{\pi}$=1$^+$ ground state and to the J$^{\pi}$=2$^+$ state at 55 keV. A J$^{\pi}$=3$^-$ assignment is proposed to the 1353 keV level on the basis of its sole decay to the J$^{\pi}$=2$^+$ state through a 1298 keV transition.  Owing to the fact that the 1481 KeV state exclusively decays to the ground state, its spin assignment could be J=0-2, providing a good candidate to the missing J$^{\pi}$=0$^+$ state.  A new level is firmly established at 1636 (2) keV from its observed three $\gamma$ decay branches 282 + 1298 + 55, 403 + 1177 and 948 + 636 +55 keV. As this level decays to the J$^{\pi}$=3$^-$, J$^{\pi}$=2$^-$ and J$^{\pi}$=3$^+$ states and not to the J=1,2 positive parity states at lower energy, it is likely to have J$^{\pi}$=4$^-$. Two levels with tentative spin J$^{\pi}$=1$^-$ and J$^{\pi}$=2$^-$ are proposed at 1749 and 1792 keV, respectively. The level at 1792 keV could as well be a candidate for the J$^{\pi}$=0$^+$ state predicted at a similar energy by the shell model calculations. A J$^{\pi}$=5$^-$ level  is proposed at 2493 keV from its 858 and 1527 keV $\gamma$ branches to the previously assigned J$^{\pi}$=4$^-$  and J$^{\pi}$=4$^+$ states at 1636 and 968 keV, respectively. Though less probable, a J$^{\pi}$=3$^-$ assignment cannot be ruled out. Other high energy $\gamma$-rays are observed at 2378, 2605, 2650 and 2874 keV. However, their placement in the level scheme is uncertain by 55 keV as there is not enough statistics to ensure that a coincidence with the 55 keV  $\gamma$-ray is present or not. Finally, the 489 keV $\gamma$-ray could not be placed without ambiguity in the level scheme as well.

\section{Discussion}
\subsection{Positive parity states along the N=17 isotones}
\begin{figure}
\centering \epsfig{width=8.5cm,file=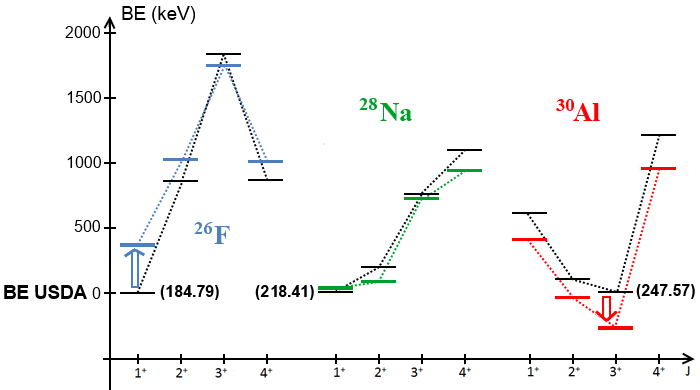}
\caption{(color on line) Comparison between the experimental binding energies of the $J^{\pi}=1^+$ - $4^+$ states in $^{30}$Al~\cite{Wang,Step} (red), $^{28}$Na~\cite{Wang,Tripa} and present work (green), and in $^{26}$F~\cite{Jura07,Sta12,Fran11,Lepailleur}  (blue) and the USDA  shell model predictions (in black, with the g.s. value written for each nucleus). The reference value for each nucleus is the calculated  binding energy of the ground state. Experimental binding energy uncertainties for the $^{30}$Al, $^{28}$Na and $^{26}$F ground states are 14, 10 and about 130 keV, respectively.}
\label{comp_SM_exp}
\end{figure}

The level scheme of $^{28}$Na shown in Fig.~\ref{schema_lvl} has been obtained by combining the results from the $\beta$-decay (left part) and the in-beam $\gamma$ spectroscopy (right part) experiments. The middle part of the spectrum displays transitions that were common to the two experiments.The two new levels with configurations J$^{\pi}=3_1^+$ and $4_1^+$ complete the quadruplet of J$^{\pi}$=1-4$^+$ states resulting from the $\pi$d$_{5/2} \otimes \nu$d$_{3/2}$ coupling in $^{28}$Na. Added to the recently discovered J$^{\pi}$=1-4$^+$ states at low energy in the N=17 isotones of $^{26}$F~\cite{Sta12, Fran11,Lepailleur} and $^{30}$Al~\cite{Step}, a systematics of their binding energies as a function of the proton-to-neutron binding energy asymmetry can be obtained and is compared to shell model calculations in Fig.~\ref{comp_SM_exp}. The energy difference between proton and neutron separation energies S$_p$-S$_n$  ranges from 15 MeV in the close-to-drip-line nucleus $^{26}$F to 7 MeV in $^{30}$Al. For each nucleus the calculated binding energy of the ground state (using the USDA interaction) is taken as the reference value for Fig.~\ref{comp_SM_exp}.

The J$^{\pi}$=1-4$^+$ states in the $^{26}$F nucleus can be described as a proton  $\pi$d$_{5/2}$  coupled to a neutron in the $\nu$d$_{3/2}$ on top of a $^{24}$O core nucleus.  In this particle-particle coupling scheme, the multiplet of states in $^{26}$F displays an upward pointing parabola in binding energy value as a function of J as seen in Fig.~\ref{comp_SM_exp}. In the present case the amplitude of the parabola scales to a first order with the strength of the residual interaction that splits the different components of the J$^{\pi}$=1-4$^+$ multiplet. With the exception of the J$^{\pi}$=3$^+$ state that is unbound and therefore may need a specific treatment, the ordering of the states in $^{26}$F is well reproduced by the USDA interaction but their calculated binding energies are too large by about 200 keV. Even if the J$^{\pi}$=1-4$^+$ states in $^{30}$Al do not have a pure configuration, they can be viewed as hole-particle coupling ($\pi$d$_{5/2})^{-1}\otimes \nu$d$_{3/2}$ states with respect to the full occupancy of the proton d$_{5/2}$ orbit. This leads to the downward pointing parabola as a function of J shown in Fig.~\ref{comp_SM_exp}.  The amplitude and shape of the parabola is extremely well reproduced using the USDA interaction. However, the calculated absolute binding energies are this time smaller than the experimental values by about 300 keV. The pattern of the binding energy of the J$^{\pi}$=1-4$^+$ states in $^{28}$Na is intermediate between $^{26}$F and $^{30}$Al owing to the fact that the d$_{5/2}$ orbit is only half-filled. The spectrum of $^{28}$Na is well reproduced by the calculations. It is worth to notice here that a similar shift in binding energy is observed along the N=17 isotones when using the USDB interaction. However such calculation proved to be less precise, predicting  the binding energy of the J$^{\pi}$=4$^+$ state wrong by about 600 keV in $^{26}$F and 400 keV in $^{30}$Al and a  J$^{\pi}$=2$^+$ ground state of $^{28}$Na instead of J=1$^+$.

Gathering all results along the N=17 isotonic chain,  a systematic deviation between experimental and theoretical binding energies is observed: as compared to experimental values the $^{26}$F is over bound, the $^{28}$Na is perfectly well reproduced, while the $^{30}$Al is under bound. A total shift in energy of about 500 keV is found between these three isotones.  Being a global shift in energy of the whole multiplet, this change in binding energy is likely carried by the monopole part of the $\pi$d$_{5/2} \otimes \nu$d$_{3/2}$ nuclear interaction rather than by multipoles.  We propose three reasons that may explain this systematic deviation. First, on the experimental side, the atomic mass of $^{26}$F derived in~\cite{Jura07}, that has been corrected in Ref.~\cite{Lepailleur} to account for the possible contamination from the 4$^+$ isomeric state, may not be correct.  This would account for the shift observed in $^{26}$F but not in $^{30}$Al. Second, the shell model interaction of~\cite{usdab} does not use an explicit isospin dependence of the Coulomb energy contribution to the binding energy of the nuclei as it should  probably be done to account for the change of nuclear radii for a wide range of isotopes. This part is therefore taken only implicitly in the fitting procedure to determine TBME from the experimental data, a feature than can cause some bias in the calculation of the binding energies if the suitable data to constrain this isospin dependent part are not available experimentally or if they are not included in the fitting procedure. Third, the scaling law of the monopole interaction in A$^{-1/3}$ may not be appropriate, and more realistic interactions should be developed for nuclei spanning over a large proton-to-neutron binding energy asymmetry.

\subsection{Negative parity states in the N=17 isotones}

In the odd-even nuclei  N=17 isotones, negative parity states 7/2$^-$ and 3/2$^-$ originating from the 0$f_{7/2}$ and 1$p_{3/2}$ orbits have been populated at about 3 MeV through (d,p) reactions: in $^{33}$S, the first 7/2$^-$ and  3/2$^-$ states lie at  2.934 and 3.220 MeV with C$^2$S values of 0.53  and 0.87, respectively~\cite{ensdf}. In $^{31}$Si, C$^2$S values for the first 7/2$^-$ and  3/2$^-$ at 3.133 and 3.535 MeV are 0.6 and 0.4, respectively~\cite{ensdf}. Recently, negative parity states have been observed in $^{29}$Mg and $^{27}$Ne. Their excitation energies are much lower, and their ordering are reversed compared to nuclei in the valley of stability.  States with L=1 and L=3  have been populated in  $^{29}$Mg at energies of 1096 and 1432 keV using the one neutron knock-out reaction from $^{30}$Mg~\cite{Terry08}.  Similar L assignments have been proposed to the states at 0.765 (3/2$^-$, L=1) and 1.74 MeV (7/2$^-$, L=3) that are populated in the $^{26}$Ne(d,p)$^{27}$Ne reaction with  C$^2$S values of 0.64(33) and 0.35(10), respectively~\cite{ne27}. Globally as protons are removed from the 0$d_{5/2}$ orbit from $^{31}$Si, the excitation energy of the two negative parity states 7/2$^-$ and  3/2$^-$ is decreasing relatively to the 3/2$^+$ ground state, and their ordering is reversed around Z=12.  It is derived that the N=20 gap between the 0$d_{3/2}$ and 0$f_{7/2}$ is collapsing and the traditional N=28 gap between the 0$f_{7/2}$ and 1$p_{3/2}$ is upside down. These features have been qualitatively attributed to the hierarchy of the proton nuclear forces in Ref.~\cite{hierarchy}. In the odd-odd nuclei such as $^{28}$Na, multiplet of negative parity states J$^{\pi}$=1$^-$-4$^-$  and J$^{\pi}$=2$^-$-5$^-$ are expected to be formed by the proton-neutron coupling 0$d_{5/2}$-1$p_{3/2}$ and 0$d_{5/2}$-0$f_{7/2}$, respectively. The present observation of tentatively assigned negative parity states J=1,2,3 and 4  in $^{28}$Na at 1749, 1233, 1353 and 1636 keV, respectively, confirms the presence of the neutron 1$p_{3/2}$ orbit at relatively low energy at Z=10 \cite{ne27}. Actually the J=1-4  negative parity states are systematically more bound than predictions by about 300 keV, hinting at a possibly stronger proton-neutron 0$d_{5/2}$-1$p_{3/2}$ interaction than calculated. The tentatively assigned J$^{\pi}$=5$^-$ state at 2496 keV (calculated at  2867 keV) is in accordance with the presence of the 0$f_{7/2}$ orbit about 1 MeV above the 1$p_{3/2}$ orbit, as proposed in $^{27}$Ne~\cite{ne27}.

\subsection{Evolution of nuclear structure toward drip-line}

In order to understand the evolution of the 1$p_{3/2}$ orbit and of the N=20 gap toward the neutron drip line, configuration-interaction (CI)  calculations have been carried out with a Hamiltonian called WBA. It is the same as the WBP Hamiltonian from \cite{wbp}, but the older USD sd-shell Hamiltonian \cite{usd} part has been replaced by the more recent USDA Hamiltonian \cite{usdab} that was used in the previous section to calculate the energy of positive-parity states. The required basis to model the N=17 isotones is the full $(1s0d)$ for positive parity states with a core of $^{16}$O. Moreover it  allows for one neutron to be excited to the $ (1p0f)  $ valence space for negative parity states.  While this basis can be used for low-lying negative-parity states in $^{25}$O, $^{26}$F or $^{27}$Ne, the dimension of the calculation is at the limit for $^{28}$Na and too large for $^{29}$Mg and $^{31}$Si.  Such calculations might be possible in the future.

The WBA interaction is used to calculate the energies of negative-parity states in  $^{27}$Ne. As they are too high by 0.4 MeV, the single-particle energies of the 1p-0f  shell orbitals have been modified by 0.4 MeV, leading to the WBA-M interaction. In Table \ref{Tab3} excitation energies of  low-lying negative parity states are given for the N=17 isotones that the model can handle. As used to constrain the  WBA interaction, the spectroscopy of  $^{27}$Ne is well reproduced. Negative parity states  are present at low energy in $^{28}$Na, as found experimentally. The wave functions of the first  J$^{\pi}$=1-4$^-$ are  composed by 60-75\% of  a neutron in the 1$p_{3/2}$ and by 20-30\% of a neutron in the  0$f_{7/2}$ orbital, the remaining weak fraction being distribution in the 1$p_{1/2}$ and 0$f_{5/2}$  orbitals. All calculated negative parity states in $^{26}$F lie above the neutron emission threshold of S$_n$ = 1.070(62)~MeV \cite{Lepailleur}, a feature that agrees with the observation of only two bound excited states of positive parity in this nucleus. In $^{25}$O, the 3/2$^-$ is expected to lie only 493 keV above the 3/2$^+$ resonance \cite{Hoff08}.  The three negative-parity states 3/2$^-$, 7/2$^-$ and 1/2$^-$ in $^{25}$O have large $^{24}$O+n spectroscopic factors. With the proximity in energy between the 3/2$^+$ and 3/2$^-$ states in $^{25}$O, $^{26}$O likely contains a significant amount of negative parity contribution in its ground and first excited states. The reliability of this extrapolation far from stability depends on the confirmation of spin assignments of negative parity states in $^{28}$Na, on the possibility to model higher Z isotones in this large valence space, as well as on the possibility to observe of negative parity states in $^{26}$F in the future.

\begin{figure}
\scalebox{0.5}{\includegraphics{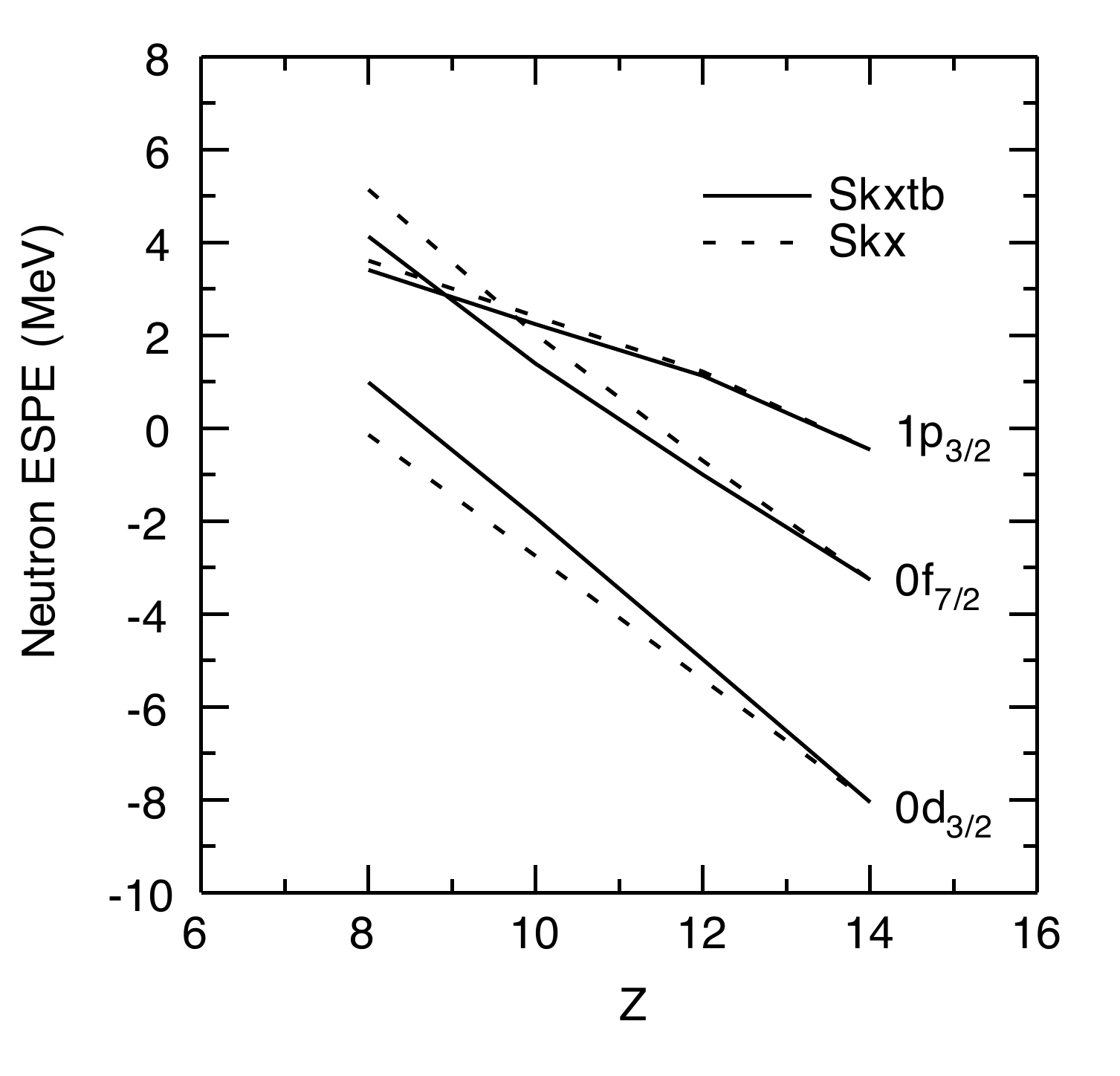}}
\caption{Effective single-particle energies
calculated for nuclei with an N=16 core using two Skyrme functionals that implements tensor forces (Skxtb) or not (Skx). The
down bending of the unbound 1p$_{3/2} $ orbital observed when reaching the drip line is due to the halo like structure of the nuclei. }
\label{ESPE}
\end{figure}

The present shell evolution towards the neutron drip line is made in an Harmonic Oscillator basis in which bound and unbound states are treated on the same footing. Therefore we look at trends of ESPE in the framework of energy density functional (EDF) calculations that do not use a Harmonic Oscillator basis. The single-particle energies obtained with the Skx \cite{skx} and Skxtb \cite{skxtb} Skyrme functionals are shown in Fig. \ref{ESPE} for an N=16 core as a function of proton number. This figure is similar to Fig.~\ref{ESPEN}, the major changes consisting here  in the down bending  of the 1p$_{3/2} $ orbital when reaching the drip line and a weaker reduction of the N=20 gap (between 0$d_{3/2}$ and 0f$_{7/2}$ orbits). This non linearity in the ESPE of the 1p$_{3/2} $ orbital is due to its halo-like nature that is reducing the effective monopole interactions there.

\begin{table}
\begin{center}
\caption{Excitation energies of some low-lying negative-parity
states obtained with the WBA-M Hamiltonian are compared to experimental values. Note that the spin assignments of the highest energy states in $^{28}$Na are tentative, see text and Fig. \ref{schema_lvl} for details. }
\label{Tab3}
\begin{tabular}{|c|c|c|c|}
\hline
Nucleus & J$^{ \pi }$ & E$_{x}$ (MeV) & E$_{x}$ (MeV) \\
        &      & theory    & experiment \\
\hline
$^{25}$O & 3/2$^{-}$ & 0.493 & \\
       & 1/2$^{-}$ & 1.898 & \\
       & 7/2$^{-}$ & 2.611 & \\
\hline
$^{26}$F & 4$^{-}$   & 1.339 & \\
       & 2$^{-}$   & 1.384 & \\
       & 1$^{-}$   & 1.952 & \\
       & 3$^{-}$   & 2.485 & \\
\hline
$^{27}$Ne & 3/2$^{-}$ & 0.825 & 0.765 \cite{Ober06,Terry06} \\
        & 7/2$^{-}$ & 1.710 & 1.74 \cite{ne27} \\
        & 1/2$^{-}$ & 1.834 & \\
\hline
$^{28}$Na & 2$^{-}$   & 1.552 & 1.233 \\
        & 3$^{-}$   & 1.715 & 1.353  \\
        & 4$^{-}$   & 1.888 &1.636 \\
        & 1$^{-}$   & 2.100 & (1.749)\\
         & 2$^{-}$   & 2.341 & (1.792)\\
          & 3$^{-}$   & 2.821  & \\
          & 0$^{-}$   & 2.836 & \\
          & 5$^{-}$   & 2.867  & 2.493\\
\hline
\end{tabular}
\end{center}
\end{table}

The ESPE for an unbound state was estimated as follows. First the central EDF potential is increased by a factor that binds the state by 0.2 MeV. Then the expectation value of the full kinetic and potential energy of this state is calculated resulting in a positive value for the ESPE that is shown. (More correctly, the properties of unbound states should be related to position and shape of neutron scattering resonances.) The weak-binding effect in the single-particle energies for the low-$\ell$ orbitals for light nuclei has been discussed by Hamamoto \cite{ham} (see Figs. 3 and 4 in \cite{ham}). It is the same mechanism that reduces the 1s$_{1/2}$-0d$_{5/2}$ gap for the region of Z=2-8 as shown by the EDF calculations in Fig. 1 of \cite{sag}, and more recently by calculations with a  Woods-Saxon potential \cite{hoff}. Basically the binding energy of the low $\ell$ orbits is more significantly affected by the proximity of continuum as they encounter a much smaller centrifugal barrier than higher $\ell$ orbits.

\begin{table}
\begin{center}
\caption{Effective single-particle energies for one
neutron outside of closed shells for $^{24}$O and $^{30}$Si.
Results are given for the configuration-interaction (CI)
and energy-density functions (EDF) models with the interactions
discussed in the text. $\Delta$ is the shell gap at N=20.}
\label{Tab4}
\begin{tabular}{|l|l|c|c|c|c|}
\hline
core  & Model & $  \epsilon (0d_{3/2})  $ & $  \epsilon (0f_{7/2})  $ & $  \epsilon (1p_{3/2})  $ &
$\Delta$ \\
      &      &    (MeV)         &     (MeV)        &    (MeV) & (MeV) \\
\hline
$^{24}$O & EDF Skx    & -0.14 & 5.14 & 3.61 & 3.75 \\
       & EDF Skxtb  & 0.99  & 4.13 & 3.41 & 2.42 \\
       & CI  WBA-M    & 0.66  & 4.88 & 2.46 & 1.54 \\
\hline
$^{30}$Si & EDF Skx    & -8.06 & -3.25 & -0.43 & 4.81 \\
        & EDF Skxtb  & -8.06 & -3.25 & -0.43 & 4.81 \\
        & CI  WBA-M    & -8.84 & -1.00 & -1.81 & 7.03 \\
\hline
\end{tabular}
\end{center}
\end{table}

Fig. \ref{ESPE} also shows the results of two different calculations. Skx does not contain a tensor interaction, while for Skxtb a tensor interaction has been added that describes the trends of single-particle energies in heavier nuclei \cite{skxtb}.
The ESPE for Skx and Skxtb are the same for Z=14 since the Skxtb tensor interaction for protons and neutrons cancel
for the doubly $  jj  $ closed shells in $^{28}$Si. From Z=14 to Z=8 protons are removed from the $  0d_{5/2}  $ orbital. The tensor interaction between the $  0d_{5/2}  $ protons and the $  0d_{3/2}  $ neutrons  increases to its maximum for Z=8. This increases the ESPE of the $  0d_{3/2}  $ orbital and makes it unbound at Z=8. It also decreases the ESPE of the $  0f_{7/2}  $ orbital. Thus, in the EDF model the N=20 shell gap at Z=8 is reduced due to a combination of the tensor-interaction effect for the $  0d_{3/2}-0f_{7/2}  $ ESPE spacing and the  weak binding effect for the $  0d_{3/2}-1p_{3/2}  $ ESPE spacing.

The ESPE have been evaluated for the WBA-M interaction at Z=8 and Z=14  by restricting the neutron core to have a $  (0d_{5/2})^{6},(1s_{1/2})^{2}  $ neutron configuration. The CI and EDF (Skxtb) results are compared  in Table \ref{Tab4}. Results are relatively close for $^{24}$O. But the CI shell gap for $^{30}$Si is larger than with EDF, explaining the earlier observation that states in  $^{28}$Na are calculated at too high energy (see Table \ref{Tab4}).

\section{Summary}
The spectroscopy of $^{28}$Na has been investigated by means of the $\beta$-decay of $^{28}$Ne at GANIL/LISE and the in-beam $\gamma$ spectroscopy through the fragmentation of $^{31,32}$Mg beams at NSCL/S800. New  positive parity states with J$^{\pi}$=3$^+$ and 4$^+$ are proposed at 691 and 968 keV, respectively, while new negative parity states are proposed at 1233, 1353, 1636  and 2493 keV with likely spin assignments J$^{\pi}$=2$^-$, J=3$^-$, 4$^-$ and 5$^-$ respectively. Other negative parity states are tentatively proposed at 1481, 1749 and 1792 keV, the spin and parity assignments of which is more uncertain. Using these complementary methods all components belonging to the multiplet of states J$^{\pi}$=1-4$^+$ arising from the proton-neutron 0d$_{5/2}$-0d$_{3/2}$ coupling have been discovered. With the recent studies of the same multiplet of states in the $^{26}$F and $^{30}$Al isotones, the evolution of the binding energy of the J$^{\pi}$=1-4$^+$ multiplet has been compared to shell model predictions using the USDA  interaction. While the relative energies of the  J$^{\pi}$=1-4$^+$ states are well reproduced with the USDA interaction in the N=17 isotones, a systematic global shift in binding energy by about 500 keV is observed when moving from the valley of stability  in $^{30}$Al to the drip line in  $^{26}$F. The origin of this change may arise from a change in the proton-neutron 0d$_{5/2}$-0d$_{3/2}$ effective interaction when exploring large proton to neutron binding energy asymmetry. Other possible reasons are proposed as well in the text.

The presence of a multiplet of negative parity states J$^{\pi}$=1-4$^-$ around 1.5 MeV, likely arising from the  $\pi$d$_{5/2}$-$\nu$p$_{3/2}$ coupling, as well as a tentative observation of a J=5$^-$ state around 2.5 MeV,  likely arising from the $\pi$d$_{5/2}$-$\nu$f$_{7/2}$ coupling, confirm the collapse of the N=20 gap and the inversion between the neutron f$_{7/2}$ and p$_{3/2}$ levels when removing protons in the d$_{5/2}$ orbital toward the drip line. These states are globally more bound than calculated by about 300 keV, a feature that may be due the high dimensionality of the basis or/and to an slightly wrong determination of the effective interactions. These features have been discussed in the framework of Shell Model and EDF calculations, leading to the conclusions that no bound negative parity state would be present in $^{26}$F and that the 3/2$^+$ ground state and 3/2$^-$ first excited states would be separated by only about 500~keV in $^{25}$O. It is important in the future to confirm the spin assignments as well as the structure of the proposed negative parity states in $^{28}$Na using for instance the $^{27}$Na(d,p)$^{28}$Na reaction.

\section{Acknowledgments}
This work is supported by the National Science Foundation (NSF) under Grant Nos. PHY-1102511, PHY-1306297 and PHY-1404442, by the OTKA contract K100835,  the German BMBF (Grant No. 05P12RDFN8) and HIC for FAIR. GRETINA was funded by the US DOE - Office of Science. Operation of the array at NSCL is supported by NSF under Cooperative Agreement PHY-1102511 (NSCL) and DOE under grant DE-AC02-05CH11231 (LBNL). F. Nowacki is acknowledged for fruitul discussions.

\end{document}